\documentclass[a4paper,11pt]{article}
\pdfoutput=1 

\usepackage{jinstpub} 

\usepackage{lineno} 

\usepackage{lipsum} 

\usepackage{xcolor}

\usepackage{subcaption}

\usepackage[final]{changes}

\definechangesauthor[name={Juan Garcia}, color=orange]{juan}
\definechangesauthor[name={Salva Ardid}, color=blue]{salva}
\definechangesauthor[name={Miquel Ardid}, color=magenta]{miquel}

\definecolor{changecolor}{rgb}{0.63,0.32,0.18}

\makeatletter

\@namedef{Changes@AuthorColor}{changecolor}
\colorlet{Changes@Color}{changecolor}

\newif\if@hide
\newcommand\hide[1]{\if@hide\else#1\fi}

\newcommand\hidemode{\@hidetrue}
\newcommand\nothidemode{\@hidefalse}
\hidemode

\makeatother

\usepackage[UKenglish,USenglish]{babel}

\title{Deep learning reconstruction in ANTARES}

\author[a]{J. García-Méndez,}
\author[b]{N. Gei\ss elbrecht,}
\author[b]{T. Eberl,}
\author[a]{M. Ardid,}
\author[a,1]{and S. Ardid\note{Corresponding author.}}

\affiliation[a]{Universitat Politècnica de València, Institut d'Investigació per a la Gestió Integrada de Zones Costaneres,\\Carrer Paranimf~1, 46730 Gandia, Spain}
\affiliation[b]{Erlangen Centre for Astroparticle Physics, Friedrich-Alexander-Universit\"at Erlangen-N\"urnberg,\\Erwin-Rommel-Str.~1, 91058 Erlangen, Germany}

\emailAdd{sardid@upv.es}

\abstract{ANTARES is currently the largest undersea neutrino telescope, located in the Mediterranean Sea and taking data since 2007. It consists of a 3D array of photo sensors, instrumenting about 10Mt of seawater to detect Cherenkov light induced by secondary particles from neutrino interactions. The event reconstruction and background discrimination is challenging and machine-learning techniques are explored to improve the performance. In this contribution, two case studies using deep convolutional neural networks are presented. In the first one, this approach is used to improve the direction reconstruction of low-energy single-line events, for which the reconstruction of the azimuth angle of the incoming neutrino is particularly difficult. We observe a promising improvement in resolution over classical reconstruction techniques and expect to at least double our sensitivity in the low-energy range, important for dark matter searches. The second study employs deep learning to reconstruct the visible energy of neutrino interactions of all flavors and for the multi-line setup of the full detector.}


\collaboration[c]{\\on behalf of the ANTARES collaboration}

\proceeding{9$^{\text{th}}$ Workshop on Very Large Volume Neutrino Telescopes\\
  18 -- 21 May 2021\\
  Valencia, Spain}

\begin{document}
\selectlanguage{USenglish}
\maketitle
\flushbottom

\section{Introduction}
\label{sec:introduction}

ANTARES, completed in 2008, is still today the largest undersea neutrino telescope and predecessor of the KM3NeT deep-sea cubic kilometre detector, which is being built. ANTARES is located 40~km off the coast of Toulon (France), at a depth of 2450~m. As neutrinos cannot be measured directly, the telescope is optimized to detect Cherenkov radiation induced by secondary particles, mainly muons, from neutrino interactions. The detector consists of 12 vertical lines spread over an area of about 0.1~km$^2$ and anchored to the seabed. Each line consists of 25 floors, with three Optical Modules (OMs) in each floor. An OM is a pressure resistant glass sphere containing a photomultiplier tube, various sensors, and the associated electronics (see ~\cite{Antares} for more information).

The reconstruction of neutrino properties, such as energy or direction, is challenging. Neutrino energies span several orders of magnitude, from tens of GeV to PeV. Bioluminescence and the decay of potassium~40 ($^{40}$K), introduces noise, as do also other environmental circumstances like sea currents. For single-line events (that is, events detected only by a single line), the reconstructions are even more complicated, due to lack of substantial information on the horizontal plane.

Reconstruction methods to estimate the direction and energy of neutrinos already exist~\cite{bbfit,energy1, energy2}, however, due to the difficulties mentioned above, their results tested on simulations are noisy and suboptimal. Our aim is using deep learning models to improve the accuracy of reconstructions.

\subsection{Approach}
\label{subsec:approach}

Deep Convolutional Networks (DCNs) have been typically applied to image classification reaching high accuracy scores. Nevertheless, over the last years they have also been applied to regression tasks reporting good estimate predictions~\cite{DCN}. We are interested in DCNs because these kind of networks are able to extract and process spatial information between image features. In this view, the ANTARES telescope is akin to a camera collecting 3D images at each time step, having as many pixels as OMs. Therefore, we transform data from each event detected by the telescope to 4D images, with time in the fourth dimension.

Regression tasks performed by DCNs estimate point predictions. To incorporate information about the quality of the reconstruction, we integrate DCNs with Mixture Density Networks (MDNs). MDNs allow us to make predictions according to a Gaussian distribution: the mean ($\mu$), which represents the most likely estimate, and its uncertainty ($\sigma$). This implies using the following loss function ($\mathcal{L}$)~\cite{MDN}:

\begin{equation}
\label{eq:L}
	\mathcal{L} = \ln (\sqrt{2\pi}\sigma^2) + \frac{1}{2}\frac{(y_{t}-\mu )^2}{\sigma^2},
\end{equation}

where $y_t$ is the true value of the property reconstructed. We train this network using Monte Carlo simulations that account for the statistics of the real data~\cite{MC}.

\section{Case studies}
\label{sec:cases}

\subsection{Reconstruction of neutrino-induced muon tracks for low-energy single-line events}
\label{subsec:tracks}

The analysis of low-energy neutrinos ($\lesssim$~100GeV) is important for studying Dark Matter. A significant proportion of events in this energy range is detected only by a single line (single-line event). Due to lack of substantial information on the horizontal plane, current methods to estimate the direction of these neutrinos only provide information about the zenith angle. We aim to improve the zenith reconstruction and to yield an estimation of the azimuth, currently missing.
In this case study, we focused on track-like events originated from charged current interactions of muon-(anti)neutrinos ($\nu_\mu^{CC}$)~\cite{CC}.

\subsubsection{Data processing}
\label{subsubsec:data1}

For single-line events, images become 2D (floor $\times$ time). The time is relative to the first photon detection (hit) of the several that represent an event. It spans a window from -200 ns to 600 ns. In each pixel of the 2D image, we insert RGB-colored content that is informative of the event reconstruction in the XY plane, which helps estimating the Azimuth angle. More specifically, we weigh the angle in which OMs are directed in XY by the recorded amplitudes in the respective OMs. From discrete hits, we create a time series with a resolution of 5 ns applying a smoothing Gaussian kernel regression to each color individually, after which we re-center the image based on the floor of the reference hit. The resulting data is then randomly split in three sets: train \added{(60\%)}, validation \added{(20\%),} and test \added{(20\%)}.

\begin{figure}[htbp]
	\centering
	\includegraphics[width=1.\textwidth]{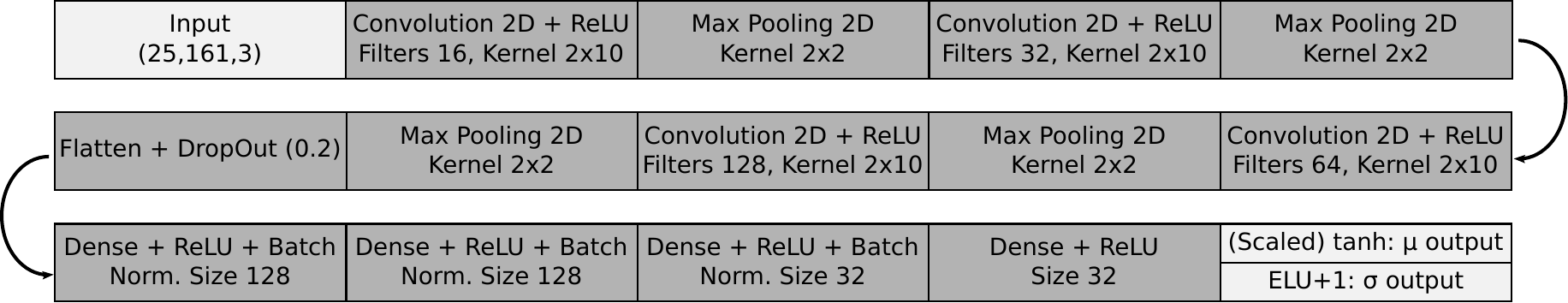}
	\caption{\label{fig:DCNtrack} \added{Details of the Deep Convolutional Network. Note that for the Zenith prediction, we scaled the tanh activation function for $\mu$, so that its value lay in [0, $\pi$] radians. No scaling was necessary for the prediction of the Azimuth since its estimation was derived from the Cartesian XY components of the unit vector.}}
\end{figure}

\added{We explored different architectures and network parameters. To diminish over-fitting during training, we considered early stopping with a patience of 10 epochs, for a maximum of 150 training epochs. In each epoch, learning batches of 64 input elements were considered. We used the Adam algorithm as the learning optimizer because it is able to dynamically regulate the learning rate, which was initialized at 0.001. Figure}~\ref{fig:DCNtrack} \added{shows the network that achieved the best results in the test set (see the results section for details):}

\subsubsection{Results}
\label{subsubsec:results1}

\begin{figure}[htbp]
	\centering
	\includegraphics[scale=.3]{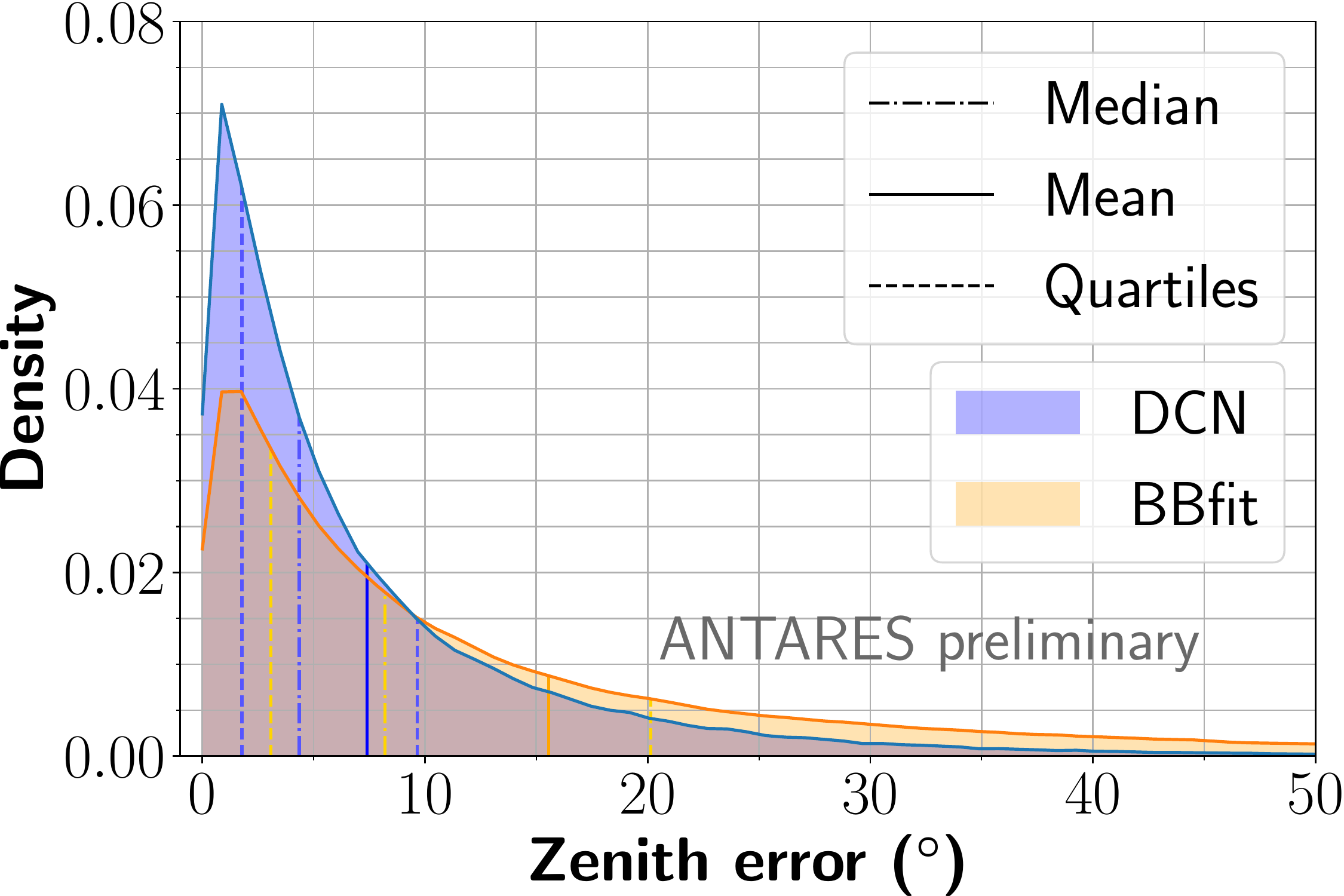}
    \includegraphics[scale=.3]{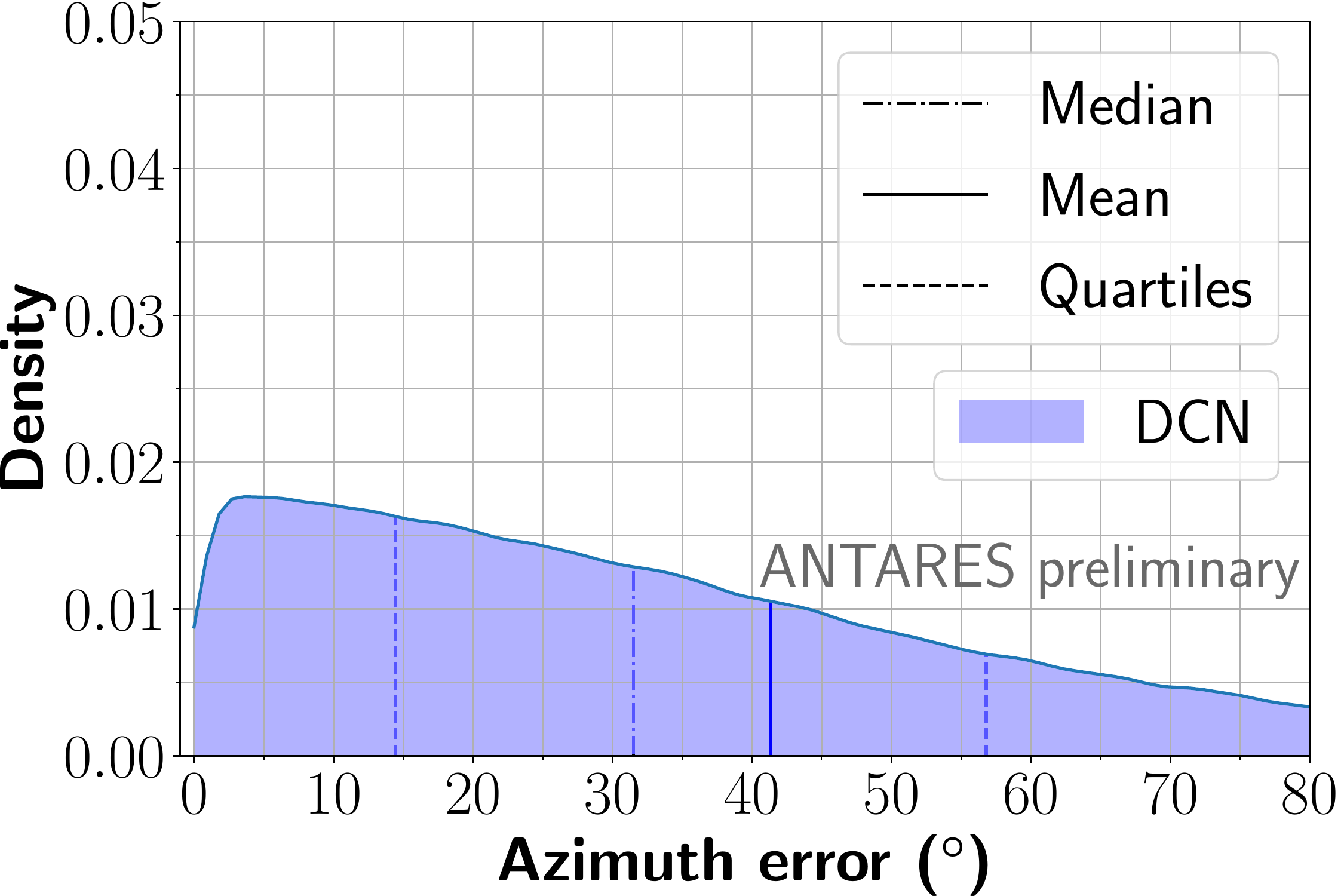}
	\caption{\label{fig:err_dist} Zenith and Azimuth error distributions.}
\end{figure}

Statistics of the error distribution for the reconstruction angles can be seen in Table~\ref{tab:results}. The error distribution is shown in Figure~\ref{fig:err_dist}. Our DCN clearly outperforms the conventional BBfit reconstruction method of ANTARES. The DCN method results in a significantly larger proportion of low errors for the Zenith angle and a first estimation for the Azimuth angle, which was previously missing for single-line events.

\begin{table}[htbp]
\centering
\caption{\label{tab:results} Mean and median absolute error of predicted neutrinos' Zenith, Azimuth and total angle.}
\smallskip
\begin{tabular}{c|c|c|c|c|c|c|}
\cline{2-7}
 & \multicolumn{2}{c|}{\textcolor{orange}{BBfit}} & \multicolumn{2}{c|}{\textcolor{blue}{DCN}} & \multicolumn{2}{c|}{\textcolor{blue}{DCN} (50\% lowest $\sigma$)} \\ \cline{2-7}
 & Mean & Median & Mean & Median & ~~Mean~~ & Median \\ \hline
\multicolumn{1}{|c|}{Zenith} & 15.5$^\circ$ & 8.5$^\circ$ & 7.4$^\circ$ & 4.4$^\circ$ & 3.7$^\circ$ & 2.5$^\circ$ \\ \hline
\multicolumn{1}{|c|}{Azimuth} & - & - & 41.4$^\circ$ & 31.5$^\circ$ & 29.3$^\circ$ & 23.6$^\circ$ \\ \hline
\multicolumn{1}{|c|}{Total}   & - & - & 28.1$^\circ$ & 22.4$^\circ$ & 18.3$^\circ$ & 13.2$^\circ$ \\ \hline
\end{tabular}
\end{table}

\begin{figure}[htbp]
	\centering
	\subfloat[Zenith (BBfit)]{{\includegraphics[scale=.3]{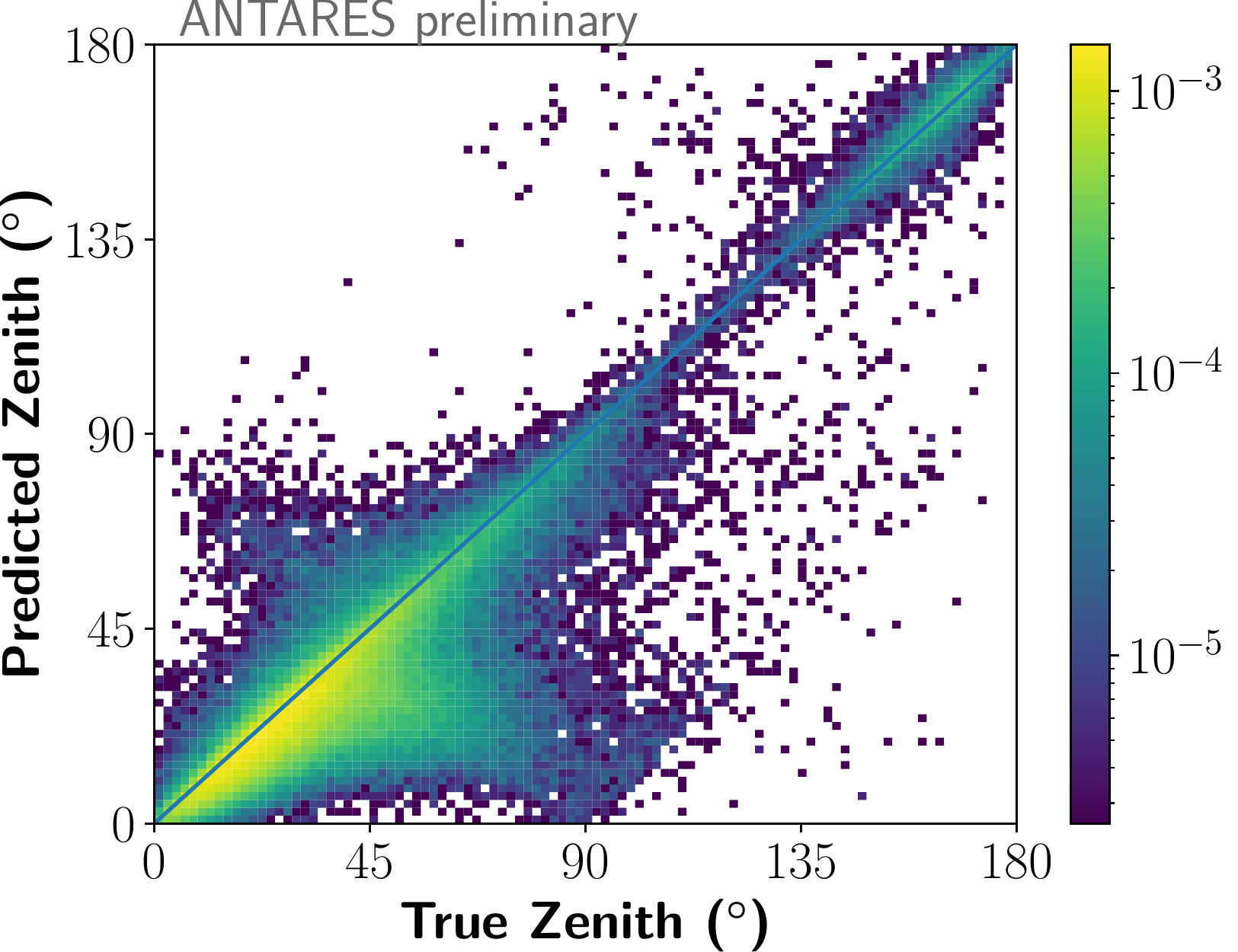}}}
	\subfloat[Zenith (DCN)]{{\includegraphics[scale=.3]{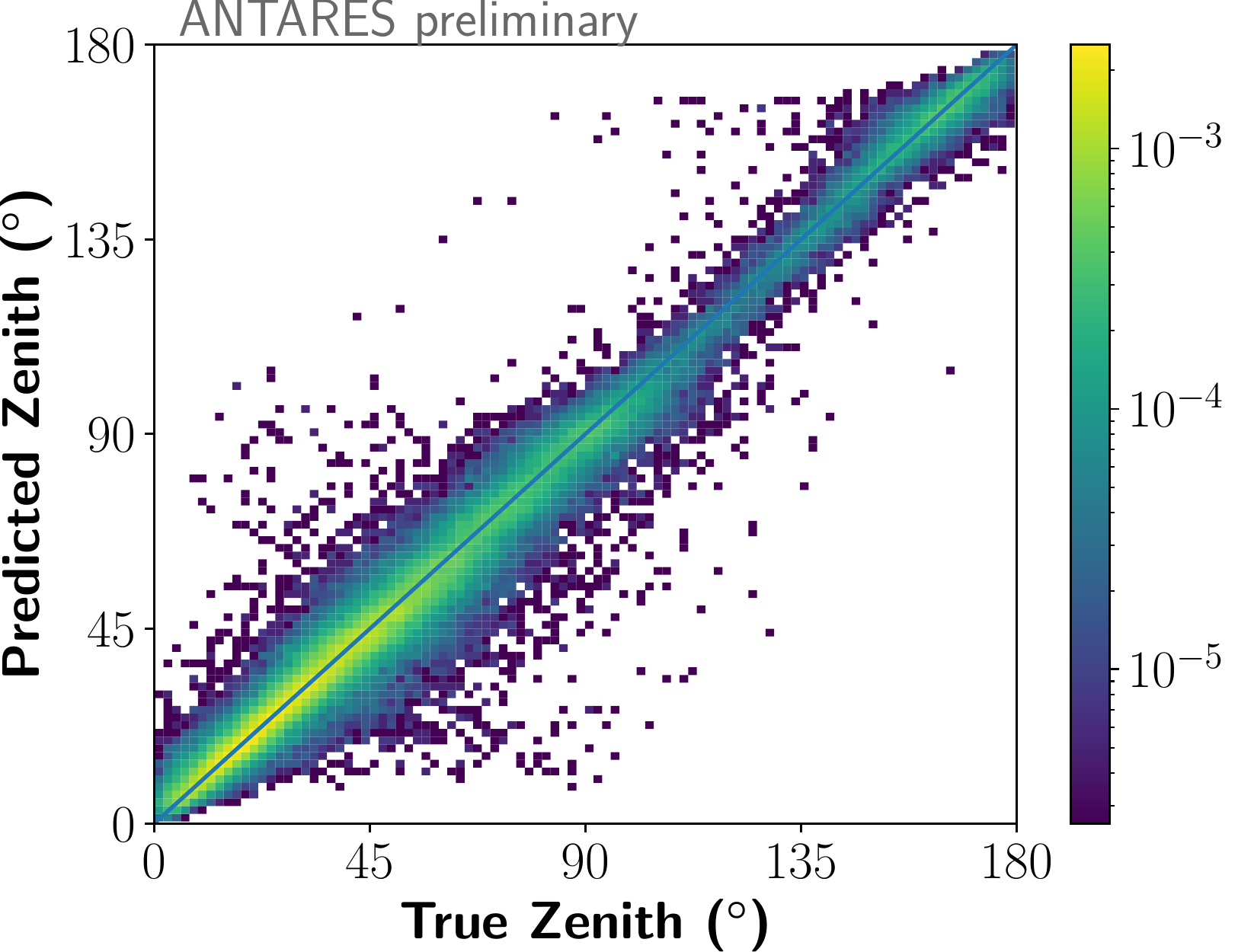}}}
    \subfloat[Azimuth (DCN)]{{\includegraphics[scale=.3]{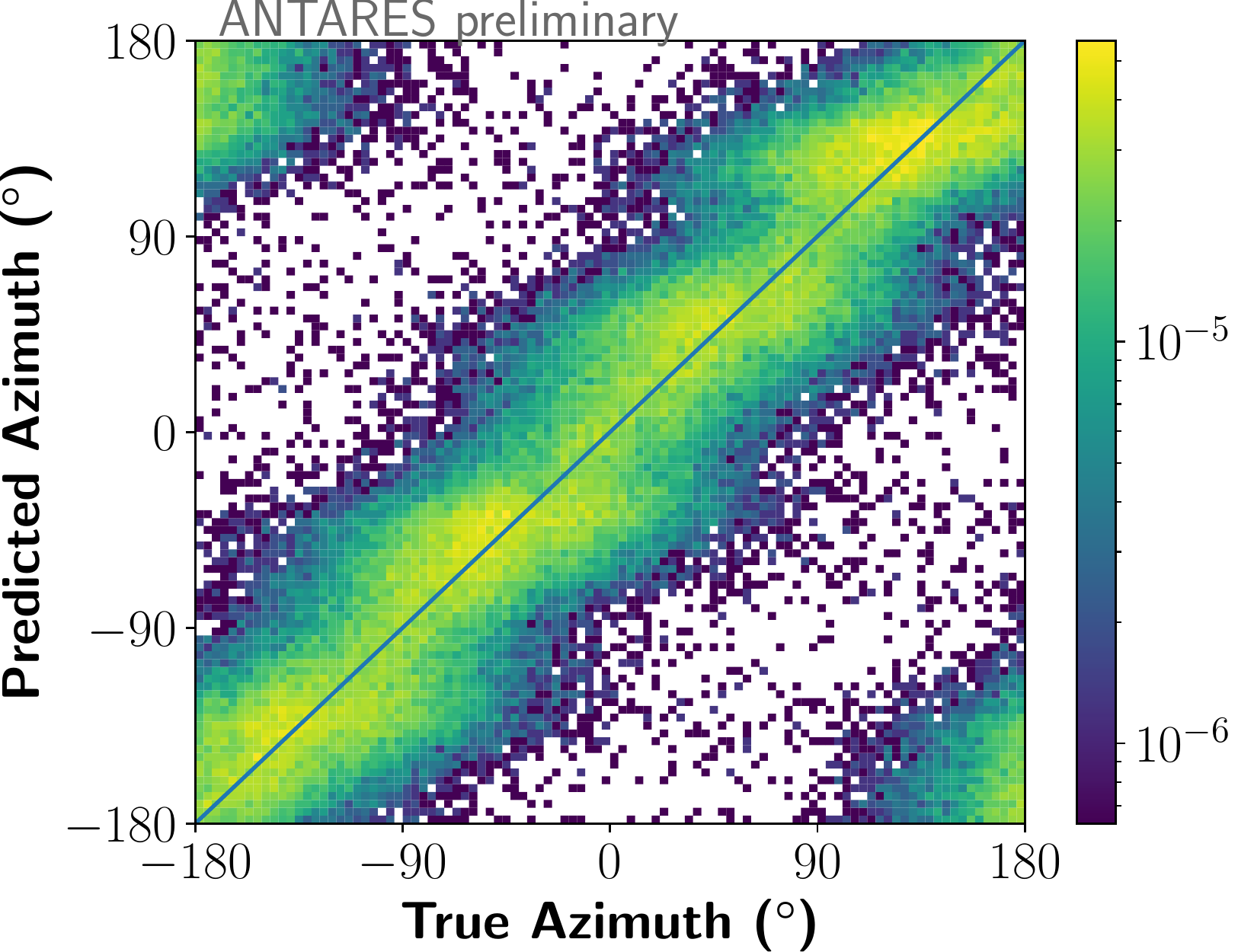}}}
	\caption{\label{fig:cut_heatmap} 2D density histograms of true angles vs. predictions for the 50\% best predictions. BBfit best predictions are taken based on its own quality parameter ($\chi^2$).}
\end{figure}

The DCN approach allows for a focus on best predictions by selecting the events with the lowest values of $\sigma$. Results are shown in Figure~\ref{fig:cut_heatmap}.

\subsection{Estimation of the neutrino energy}
\label{subsec:energy}

The energy reconstruction is designed for both track- and shower-like multi-line neutrino events. The main approach is based on the methods developed for KM3NeT/ORCA which have shown promising results \cite{CNNKM3NeT}. The network is trained on charged and neutral current interactions of electron and muon (anti)neutrinos. The training set is balanced equally between event types but no further balancing regarding the energy is applied. The goal of this reconstruction is to predict the visible energy. The visible energy is directly related to the actual neutrino energy for charged current events. In case of neutral current interactions, the energy that is transferred to the hadronic system constitutes the visible energy. For a detailed description of the energy reconstruction, see \cite{thesis}.

\subsubsection{Data processing}
\label{subsubsec:data2}

In the case of reconstructing multi-line events, the DCN approach uses four dimensional input images containing the full spatial and temporal hit information. Therefore, the spatial binning is chosen such that every floor is in one pixel of the image, leading to a pixel grid of $4 \times 4 \times 25$. The response of the three OMs on one floor is summed up. The number of time bins is set to 100 after applying an asymmetric cut of 2000\,ns around the time of the first triggered hit. This leads to a time resolution of 20\,ns. Hence, the dimensions of the input images is $4 \times 4 \times 25 \times 100$.

\subsubsection{Results}
\label{subsubsec:results2}

\begin{figure}[htbp]
	\centering
	\includegraphics[width=.4\textwidth]{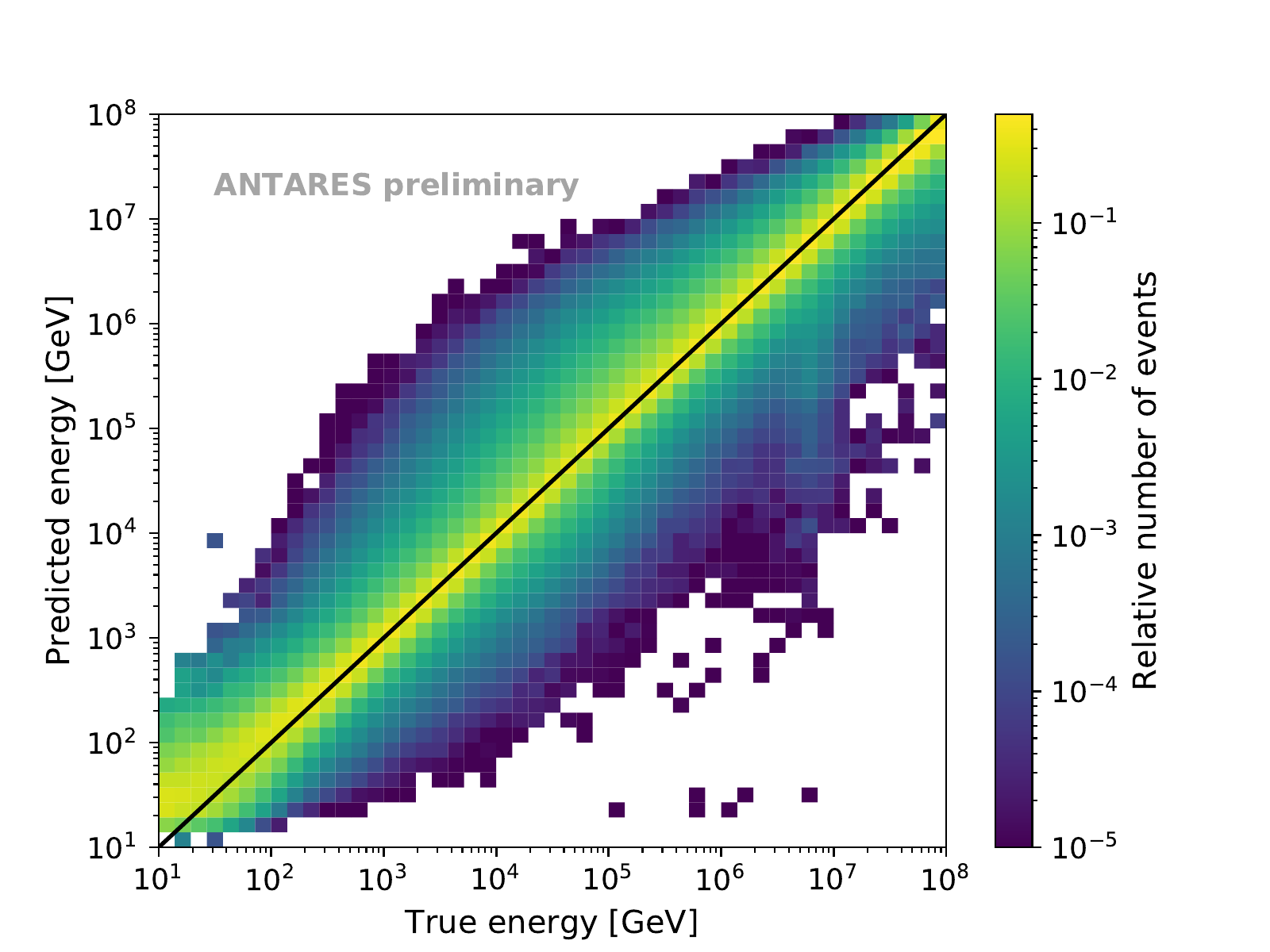}
	\includegraphics[width=.4\textwidth]{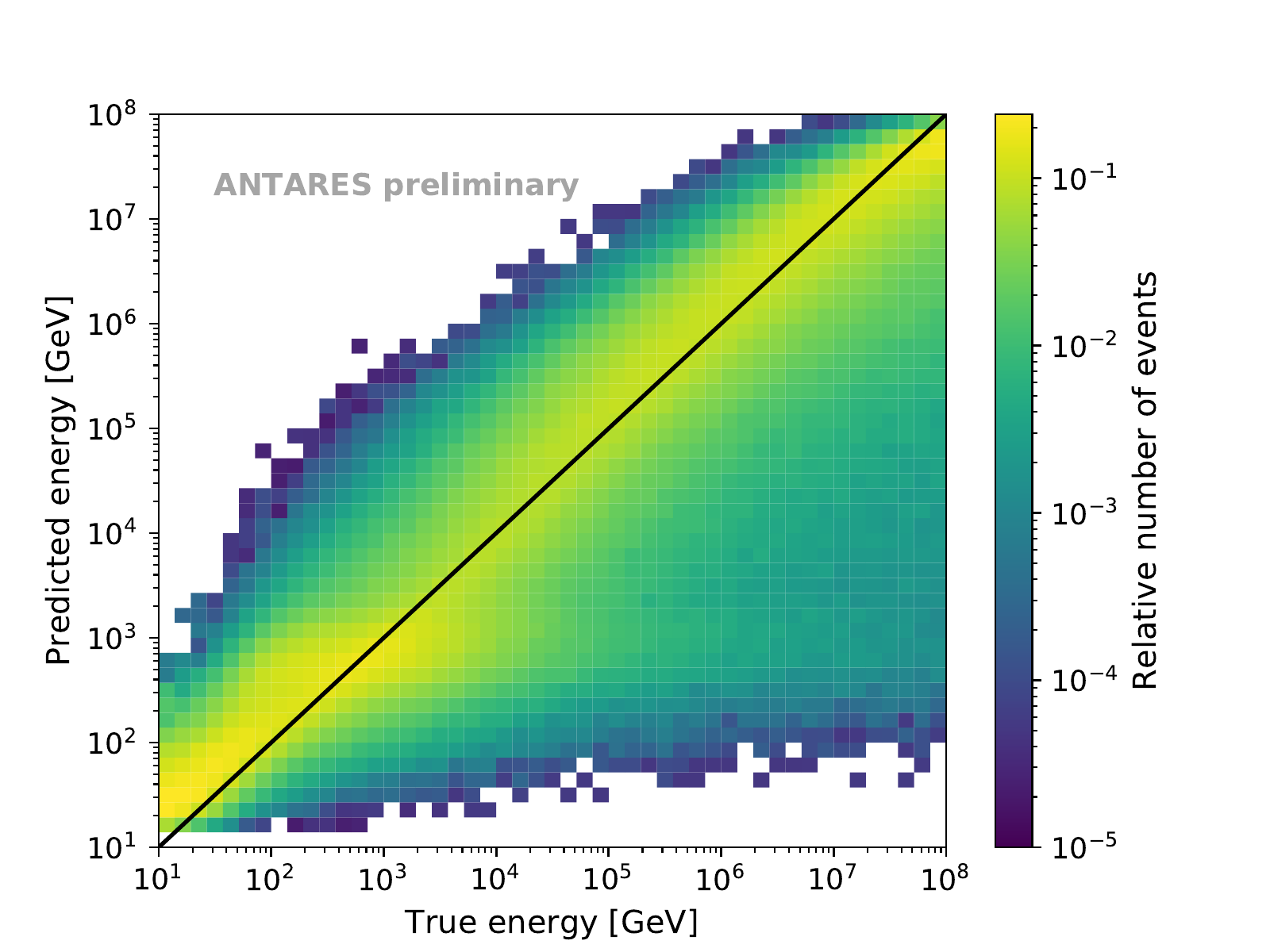}
	\caption{\label{fig:true_pred} Event distributions as a function of true vs predicted energy for shower-like (left) and track-like (right) neutrino events. The distributions have been normalized column-wise. Bin values lower than $10^{-5}$ are not shown.}
\end{figure}

Figure \ref{fig:true_pred} presents the event distributions for predicted versus true energy, separately for track- and shower-like neutrino events. As it can be seen, the performance differs for both event types.

In case of shower-like events, the correlation is linear over a large energy range with a resolution between 0.15 and 0.25 for true energies above 1\,TeV. However, there is an excess of events with underestimated energy around a true energy of a few PeV. This is the result of events with a certain amount of invisible energy from the Glashow resonance.

Considering the track-like events, the dominant observation is a significant amount of events with underestimated energy, especially at larger energies. This is caused by events that are not fully contained inside the instrumented volume of ANTARES. On the other hand, the predicted energy is slightly too high for lower energetic events. The transition of these two regions is around 1\,TeV which is about the muon critical energy in water.

In summary, the performance of the network seems to be determined by physics-related factors such as different light emission characteristics above and below the critical energy as well as statistical fluctuations caused by the unbalanced training set regarding the energy. Also in the DCN approach the uncertainty $\sigma$ on the predicted energy is reconstructed and is intended to work as a cut parameter for the rejection of poor predictions, cf. Sec.~1.1. It can be shown that more conservative cuts primarily reduce overestimation. Additionally, the survival rate for a given cut on the uncertainty for shower-like events is much better than in the case of track-like events. In the case of a neutrino spectrum $\sim E^{-2}$, a cut that preserves around 75\,\% of all shower-like events leads to a survival rate of only 10\,\% for track-like events.

\section{Discussion}
\label{sec:discussion}

Our results show that DCNs are a good tool for event reconstruction in ANTARES. For single-line track reconstruction, we achieved improved Zenith estimations and an estimation of the Azimuth angle that was previously missing. The estimation of the neutrino energy shows promising results as well, especially for shower-like events. Additionally, the DCN approach provides better quality parameters than those from currently applied conventional methods both for direction and energy reconstructions. Next step is applying these methods to real data from ANTARES.

\acknowledgments

The authors acknowledge the financial support of the Generalitat Valenciana Gen--T Program (ref. CIDEGENT/2019/043) and Ministerio de Ciencia e Innovación/European Union (FEDER): Programa Estatal de Generación de Conocimiento (ref. PGC2018-096663-B-C43).






%
%
%

\end{document}